\documentclass[journal=jacsat,manuscript=article]{achemso}

\usepackage{chemformula} 
\usepackage[T1]{fontenc} 
\usepackage{xcolor}
\usepackage{indentfirst}

\author{Laith Alahmed}
\affiliation[Auburn University]
{Department of Electrical and Computer Engineering, Auburn University, Auburn, AL 36849}
\author{Bhuwan Nepal}
\affiliation[University of Alabama]
{Department of Physics, University of Alabama, Tuscaloosa, AL 35487}
\author{Juan Macy}
\affiliation[Florida State University]
{National High Magnetic Field Laboratory, Florida State University, Tallahassee, Florida 32310}
\author{Wenkai Zheng}
\affiliation[Florida State University]
{National High Magnetic Field Laboratory, Florida State University, Tallahassee, Florida 32310}
\author{Arjun Sapkota}
\affiliation[University of Alabama]
{Department of Physics, University of Alabama, Tuscaloosa, AL 35487}
\author{Nicholas Jones}
\affiliation[Auburn University]
{Department of Electrical and Computer Engineering, Auburn University, Auburn, AL 36849}
\author{Alessandro R. Mazza}
\affiliation[Oak Ridge National Laboratory]
{Materials Science and Technology Division, Oak Ridge National Laboratory, Oak Ridge, TN, 37831, USA}
\author{Matthew Brahlek}
\affiliation[Oak Ridge National Laboratory]
{Materials Science and Technology Division, Oak Ridge National Laboratory, Oak Ridge, TN, 37831, USA}
\author{Wencan Jin}
\affiliation[Auburn University]
{Department of Physics, Auburn University, Auburn, AL 36849}
\author{Masoud Mahjouri-Samani}
\affiliation[Auburn University]
{Department of Electrical and Computer Engineering, Auburn University, Auburn, AL 36849}
\author{Steven S.L. Zhang}
\affiliation[Case Western Reserve University]
{Department of Physics, Case Western Reserve University, Cleveland, Ohio 44106}
\author{Claudia Mewes}
\affiliation[University of Alabama]
{Department of Physics, University of Alabama, Tuscaloosa, AL 35487}
\author{Luis Balicas}
\affiliation[Florida State University]
{National High Magnetic Field Laboratory, Florida State University, Tallahassee, Florida 32310}
\author{Tim Mewes}
\affiliation[University of Alabama]
{Department of Physics, University of Alabama, Tuscaloosa, AL 35487}
\author{Peng Li}
\affiliation[Auburn University]
{Department of Electrical and Computer Engineering, Auburn University, Auburn, AL 36849}
\email{tmewes@ua.edu,balicas@magnet.fsu.edu,wzj0029@auburn.edu,pzl0047@auburn.edu}

\title[An \textsf{achemso} demo]
  {Magnetism and Spin Dynamics in Room-Temperature van der Waals Magnet Fe$_5$GeTe$_2$}

\abbreviations{IR,NMR,UV}

\begin{document}

%
%
%
%
%
\begin{abstract}

Two-dimensional (2D) van der Waals (vdWs) materials have gathered a lot of attention recently. However, the majority of these materials have Curie temperatures that are well below room temperature, making it challenging to incorporate them into device applications. In this work, we synthesized a room-temperature 2D magnetic crystal Fe$_5$GeTe$_2$ with a Curie temperature \textit{T}$_c = 332$ K, and studied its magnetic properties by vibrating sample magnetometry (VSM) and broadband ferromagnetic resonance (FMR) spectroscopy. The experiments were performed with external magnetic fields applied along the \textit{c}-axis (\textit{H}$\parallel$\textit{c}) and the \textit{ab}-plane (\textit{H}$\parallel$\textit{ab}), with temperatures ranging from 300 K to 10 K. We have found a sizable Landé \textit{g}-factor difference between the \textit{H}$\parallel$\textit{c} and \textit{H}$\parallel$\textit{ab} cases. In both cases, the Landé \textit{g}-factor values deviated from \textit{g} = 2. This indicates a possible contribution of orbital angular momentum to the magnetic moment but may also be caused by lack of saturation at FMR. The FMR measurements also show a broadened linewidth at low temperatures; together with the VSM data, our measurements indicate that magnetic domains of different orientations develop within Fe$_5$GeTe$_2$ in both \textit{H}$\parallel$\textit{ab} and \textit{H}$\parallel$\textit{c} cases, with a higher level of randomness in the \textit{H}$\parallel$\textit{c} case, especially at lower temperatures. Our experiments highlight key information regarding the magnetic state and spin scattering processes in Fe$_5$GeTe$_2$, which promote the understanding of magnetism in Fe$_5$GeTe$_2$, leading to implementations of Fe$_5$GeTe$_2$ based room-temperature spintronic devices.

\end{abstract}

\section{I. Introduction}

The increasing interest in magnetic two-dimensional (2D) van der Waals (vdWs) materials in recent years is warranted by their importance for fundamental studies of 2D magnetism, as well as potential applications for spintronic devices. Compared to three-dimensional (3D) magnets, 2D magnetic materials exhibit exotic electrotransport, optical, and spin properties.\cite{Frisenda2020,Mak2019,mitta_2dmaterials} One of the biggest practical issues of most 2D vdWs magnetic materials is that they generally have a Curie temperature (\textit{T}$_\textrm{c}$) that is well below room temperature, making it difficult to incorporate them into relevant devices.\cite{doi:10.1063/1.4914134,Carteaux_1995,doi:10.1143/JPSJ.15.1664,doi:10.1021/cm504242t,doi:10.1002/ejic.200501020,doi:10.7566/JPSJ.82.124711} For example, the Curie temperatures of 2D magnetic materials such as Cr(Si,Ge)Te$_3$ (33 K and 61 K)\cite{doi:10.1063/1.4914134,Carteaux_1995}, Cr(Br,I)$_3$ (47 K and 61 K)\cite{doi:10.1143/JPSJ.15.1664,doi:10.1021/cm504242t}, and Fe$_3$GeTe$_2$ (220 K)\cite{doi:10.1002/ejic.200501020,doi:10.7566/JPSJ.82.124711}, are all significantly below 300 K. This is due to their 2D nature, where the pair-exchange interaction is much weaker than in 3D magnets, as it is mostly mediated by neighboring magnetic atoms in the 2D plane. The interactions between the magnetic layers across the vdWs gaps are much weaker in comparison. 

The low \textit{T}$_\textrm{c}$ of the aforementioned 2D vdWs ferromagnets makes it impossible to use them in  room-temperature spintronic devices. Furthermore, a potential challenge arises when approaching the 2D limit of vdWs crystals, explained by the Mermin-Wagner theorem \cite{PhysRevLett.17.1133}, that intrinsic long-range magnetic order cannot be observed in the isotropic Heisenberg magnet due to strong thermal fluctuations. This hinders the emergence of long-range magnetic order. However, uniaxial magnetic anisotropy may come to the rescue by stabilizing the ferromagnetic order against finite thermal fluctuations. More specifically, in the 2D limit, it was shown theoretically that the Curie temperature is given by the uniaxial magnetic anisotropy constant \textit{K}, and the spin-exchange interaction \textit{J}, as follows\cite{PhysRevB.38.12015}: 
\begin{equation}
    T_c \sim \frac{4\pi J}{3ln(\pi^2J/K)}\label{critical_temp}
\end{equation}
As the magnetic anisotropy in vdWs ferromagnets, set by the spin-orbit coupling, is much smaller than the exchange interaction,  \textit{T}$_c$ is low.\cite{Seoeaay8912} Extensive research efforts succeeded in engineering materials that could overcome these challenges. For example, \textit{T}$_c$ can be significantly raised to about room temperature by enhancing exchange interaction while keeping the vdWs structure,\cite{Seoeaay8912} such as in the layered 2D Fe$_n$GeTe$_2$ (\textit{n}$\geq$ 3)\cite{Deng2018,Li2020v2}. This led to Fe$_3$GeTe$_2$ with \textit{T}$_c$ around 220 K\cite{PhysRevB.93.144404,doi:10.1021/acs.inorgchem.5b01260,doi:10.7566/JPSJ.82.124711}, Fe$_4$GeTe$_2$ with \textit{T}$_c$ = 270 K\cite{Seoeaay8912}, and Fe$_{5}$GeTe$_{2}$ with T$_\textrm{c}$ ranging from 260 - 310K, depending on the Fe content\cite{Li2020v2,doi:10.1021/acsnano.8b09660,PhysRevMaterials.3.104401}.

Ferromagnetic resonance (FMR) spectroscopy is an important technique to study magnetization dynamics \cite{Oates2002,doi:10.1063/1.4852415}. Several studies have analyzed the spin dynamics in 2D magnets by FMR \cite{PhysRevLett.124.017201,PhysRevMaterials.4.064406,PhysRevB.100.134437}, which revealed their magnetocrystalline anisotropy dependence on temperature. It is found that there is a discrepancy between the Landé \textit{g}-factor along the \textit{c}-axis and the \textit{ab}-plane directions in the 2D magnet Cr$_2$Ge$_2$Te$_6$ \cite{PhysRevB.100.134437,FGTAPL}. However, the Landé \textit{g}-factor is quite isotropic along different directions in another 2D magnet CrI$_3$ \cite{PhysRevMaterials.4.064406}. These measurements were all performed on 2D magnets with low Curie temperatures (e.g. $T_\textrm{c}$ = 61 K for CrI$_\textrm{3}$). With the most promising room-temperature vdWs magnet arguably being Fe$_5$GeTe$_2$, we are interested in understanding its quasi-statis and dynamic magnetic properties, which can shed light on its magnetic states and spin scattering mechanisms.


We first synthesize the vdWs magnet Fe$_5$GeTe$_2$, and then we study its magnetization properties using both vibrating sample magnetometry (VSM) and ferromagnetic resonance (FMR) spectroscopy, in the temperature range of 300 K to 10 K. For FMR, a microwave field was applied to the sample in addition to a quasistatic magnetic field, thus triggering spin precession. At the resonance field \textit{H}$_\textrm{res}$ for a given microwave frequency \textit{f}, FMR oscillation (uniform-mode excitations of spin waves with \textit{k} $\approx$ 0) occurs. The FMR spectroscopy has revealed different Landé \textit{g}-factors along the \textit{c}-axis and the \textit{ab}-plane in Fe$_5$GeTe$_2$, indicative of possible orbital momentum contribution to the magnetic moment. After examining the temperature dependence of the FMR linewidth, we conclude that Fe$_5$GeTe$_2$ maintains a ferromagnetic phase with randomized magnetic domains at lower temperatures.


\section{II. Structure Characterization}
Nominal $\mathrm{Fe_5GeTe_{2}}$  crystals are grown using a mixture of precursor materials filled into a quartz ampoule that is vacuumed and sealed with 1.9 mg/cm$^{3}$ of iodine as a transport agent. The mixture consists of pure elements of Fe:Ge:Te in the molar ratio of 6.2:1:2 (Fe: $99.998\%$, powder, Alfa Aesar; Ge: $99.999\%$, 100 mesh, Alfa Aesar; Te: $99.999\%$, powder, Alfa Aesar). The excess Fe powder is to compensate for any possible Fe-site vacancies that might occur during the growth.
        
A standard MTI 2-zone model OTF-1200X furnace was employed, where the reactants or elemental precursors were placed in the high-temperature zone and the products were grown in the low-temperature side. The ramping rate for both the hot (775 $^{\circ}$C) and cold (700 $^{\circ}$C) zones to their target temperatures was 1 $^{\circ}$C/min. This temperature differential was held for 14 days with the $\mathrm{Fe_5GeTe_{2}}$ crystals being subsequently quenched in an ice bath. 
        
Prior to characterization, the excess iodine was removed through a bath and rinse cycle of acetone and isopropyl alcohol, respectively. Samples were either stored in a glove box with high purity argon gas ($99.99\%$) of 0.01 ppm O$_{2}$/H$_{2}$O, or a desiccator under vacuum with pressures ranging between 100-200 mTorr.

\begin{figure}[h]
    \centering
    \includegraphics[width=1\textwidth]{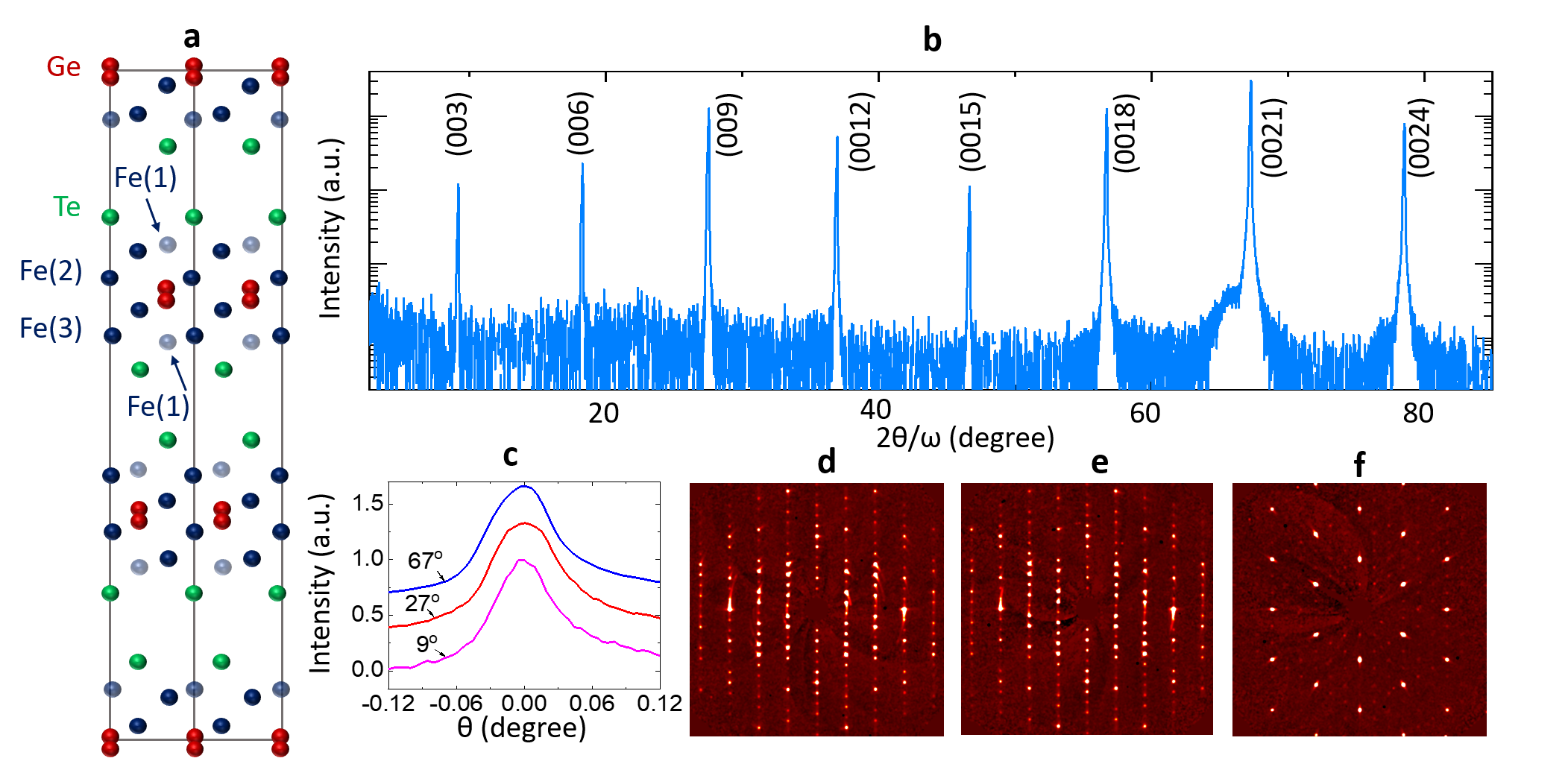}
    \caption{Crystal structure and x-ray diffraction (XRD) of single crystal Fe$_5$GeTe$_2$. \textbf{a}. Schematic of crystal structure of Fe$_5$GeTe$_2$. \textbf{b}. XRD 2$\theta$/$\omega$ scan showing (00\textit{l}) peaks. \textbf{c}. Rocking curve scan of the peaks at 9$^\circ$, 27$^\circ$, 67$^\circ$ showing high crystallinity. \textbf{d-f}: Single crystal XRD scan of Bragg reflections of different planes. \textbf{d}: (0\textit{kl}) plane. \textbf{e}: (\textit{h}0\textit{l}) plane. \textbf{f}: (\textit{hk}0) plane.}
    \label{fig:fig1}
\end{figure}

The crystalline structure characterization is measured on a $\mathrm{Fe_5GeTe_{2}}$ bulk crystal with properties shown in Figure 1. Figure \ref{fig:fig1}a shows the crystal structure schematic of Fe$_5$GeTe$_2$. The vdW-separated eight atomic-thick monolayers of two unit cells can be observed, where the vdWs gaps exist between the Te atoms of neighboring unit cells. The light-blue circles labeled Fe(1) represent the two possible occupation locations for the Fe(1) atoms, either above or below a given Ge atom, with an occupation probability not exceeding 50\%, as the Fe-Ge bond becomes non-physical if both locations are occupied simultaneously\cite{doi:10.1021/acsnano.8b09660}. 

The X-ray diffraction (XRD) data collected for the experimental Fe$_5$GeTe$_2$ sample crystal are shown in Figure \ref{fig:fig1}b. The (00\textit{l}) reflections reveal the $c$-axis of the single crystal. The (00\textit{l}) peaks, where $l$=3$n$, reflect an \textit{ABC} stacking sequence in the unit cell of the bulk crystal. This is consistent with the rhombohedral lattice structure of the R3m (No. 166) space group, as previously reported\cite{doi:10.1021/acsnano.8b09660,Li2020v2}.

The rocking curves measured at (00\textit{l}) peak angles are shown in Figure \ref{fig:fig1}c. The full-width-at-half-maximum values of the acquired curves, with values less than 0.06$^\textrm{o}$, reflect the high level of crystallinity of the Fe$_5$GeTe$_2$ samples. Figures 1d-f are Bragg reflection scans of different crystal planes (0\textit{kl}), (\textit{h}0\textit{l}), (\textit{hk}0) from high-resolution XRD. They all show clear streaks, confirming the high-quality of the single-crystal samples.

\section{III. Quasi-Static Magnetization Properties}
Quasi-static magnetization properties were measured using VSM in a Quantum Design Dynacool PPMS system. The measurements were carried out with the magnetic field applied along both the \textit{c}-axis (\textit{H}$\parallel$\textit{c}) and the \textit{ab}-plane (\textit{H}$\parallel$\textit{ab}) directions. To determine the Curie temperature of the sample, we measured field-cooled curves, as well as heat capacity curves in the absence of a magnetic field.

\begin{figure}[h]
    \centering
    \includegraphics[width=0.85\textwidth]{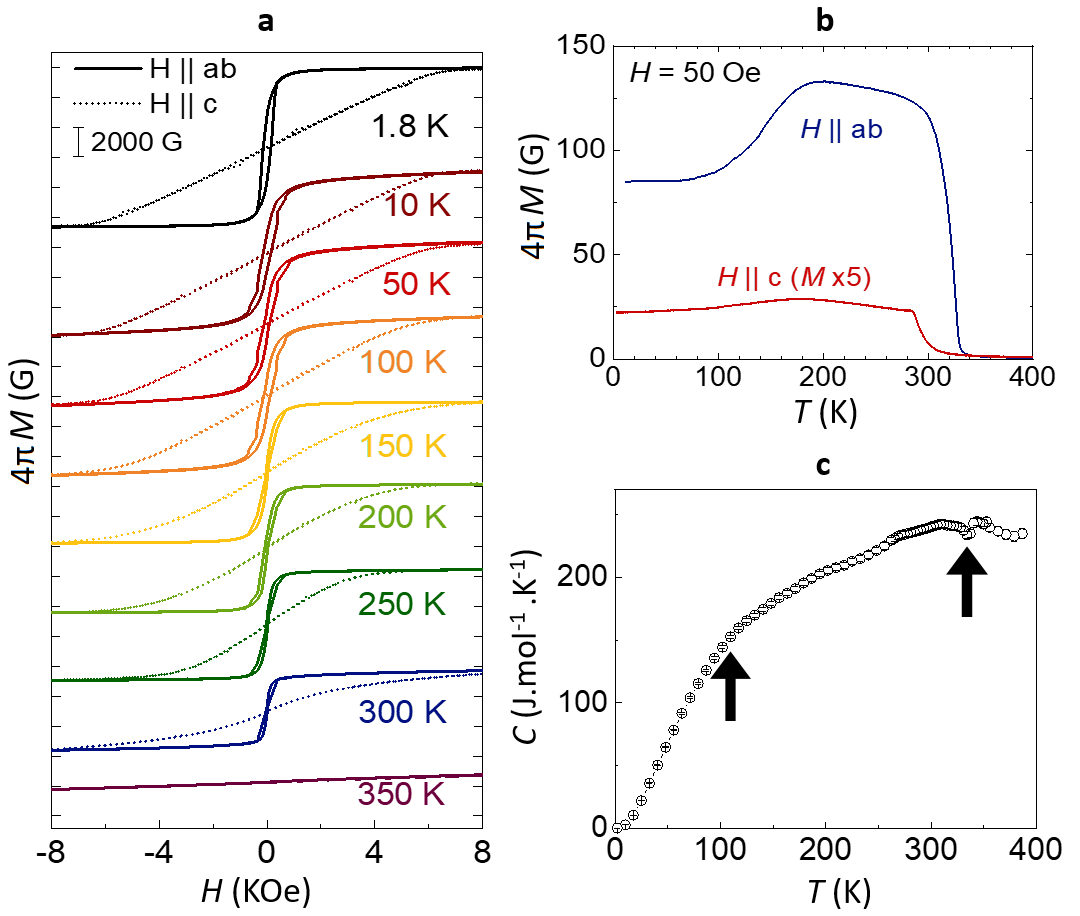}
    \caption{Static magnetization of the Fe$_5$GeTe$_2$ bulk single crystal. \textbf{a}. Temperature-dependent hysteresis loops at various temperatures for \textit{H}$\parallel$\textit{c} (dashed lines) and \textit{H}$\parallel$\textit{ab} (solid lines). \textbf{b}. Field-cooled curves (\textit{H} = 50 Oe) for the \textit{H}$\parallel$\textit{c} and the \textit{H}$\parallel$\textit{ab} cases, respectively. \textbf{c}. Heat capacity as a function of temperature. The transition at 332 K and 110 K mark the Curie temperature and possible magnetic phase transition, respectively.}
    \label{fig:fig2}
\end{figure}

Figure \ref{fig:fig2}a shows the results of the VSM magnetization versus field measurements of the Fe$_5$GeTe$_2$ sample, for temperatures ranging from 1.8 K to 350 K, and for \textit{H}$\parallel$\textit{c} (dashed lines) and \textit{H}$\parallel$\textit{ab} (solid lines). The curves show an in-plane anisotropy due to strong demagnetizing field along the \textit{c}-axis. Possible spin-reorientation features, such as the ones observed in Fe$_4$GeTe$_2$ \cite{Seoeaay8912}, are not observed in this sample. Our results are reasonable considering the fact that the out-of-plane magnetocrystalline anisotropy in Fe$_5$GeTe$_2$ crystals is very weak, i.e., 0.39 J/cm$^3$ \cite{Seoeaay8912, PhysRevB.102.064417}. Figure \ref{fig:fig2}b shows the field-cooled (FC) curves for the \textit{H}$\parallel$\textit{c} and the \textit{H}$\parallel$\textit{ab} cases. The magnetization magnitude change between 100 K and 200 K in the \textit{H}$\parallel$\textit{ab} curve indicates possible magnetic phase transitions. This feature has been reported in previous publications \cite{Seoeaay8912,PhysRevB.102.064417,doi:10.1021/acsnano.8b09660}. The Curie temperature of the sample is determined to be \textit{T}$_\textrm{c}$ = 332 K according to the transition points in the FC curves. 

Heat capacity measurements were used to validate the Curie temperature estimation. The measurements were set to start from the highest temperature setpoint, \textit{T} = 390 K, then the temperature was gradually reduced to 1.8 K as the heat capacity data was collected. This procedure guarantees an appropriate time constant is used to achieve more stable readings. The measurement results are shown in Figure \ref{fig:fig2}c. Two points of interest are highlighted on the curve by two black arrows. The first is a transition at \textit{T} = 332 K, which is consistent with \textit{T}$_\textrm{c}$ from the FC measurement. The second is the observation of a slope change around \textit{T} = 110 K. The slope change indicates that Fe$_5$GeTe$_2$ experiences some phase transition, and we will discuss it more extensively in the analysis of FMR linewidth in Section IV.

\section{IV. Magnetization Dynamics}
We measured the FMR response of the Fe$_5$GeTe$_2$ sample for \textit{H}$\parallel$\textit{c} and for \textit{H}$\parallel$\textit{ab}, at temperatures varying from 10 K to 300 K. In our custom-built system, a coplanar waveguide (CPW) with impedance matched to 50 $\Omega$ was used to guide the microwave field to the sample. The tested microwave frequencies ranged from 5 GHz to 40 GHz. For each microwave frequency, the magnetic field was swept from 15 kOe to zero. A microwave diode was used to convert the transmitted microwave signal to a dc voltage. To improve the signal-to-noise ratio, we used a set of field-modulation coils supplemented by a lock-in amplifier to detect the signal. Thus, the detected FMR response is identified as the derivative of the microwave power absorption.

\begin{figure}[t]
    \centering
    \includegraphics[width=0.8\textwidth]{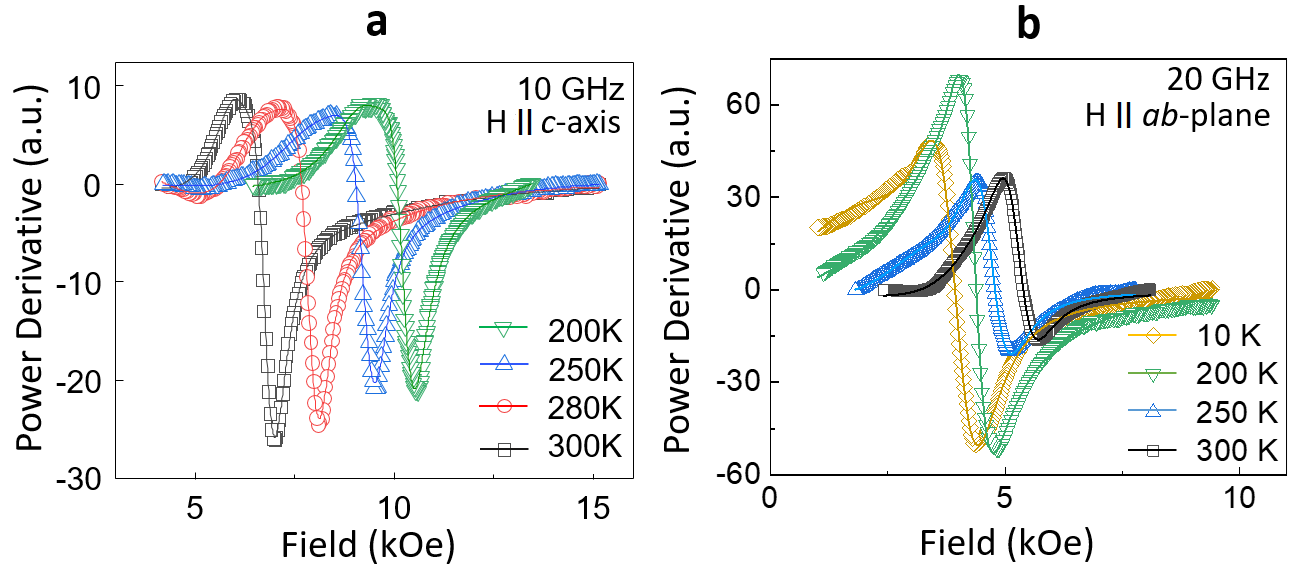}
    \caption{Ferromagnetic resonance (FMR) measurements of Fe$_5$GeTe$_2$ single crystal. \textbf{a}. FMR profiles for \textit{H}$\parallel$\textit{c} at 200 K, 250 K, 280 K, and 300 K.  \textbf{b}. FMR profiles for \textit{H}$\parallel$\textit{ab} at 10 K, 200 K, 250 K and 300 K.}
    \label{fig:fig3}
\end{figure}

\begin{figure}[h]
    \centering
    \includegraphics[width=0.8\textwidth]{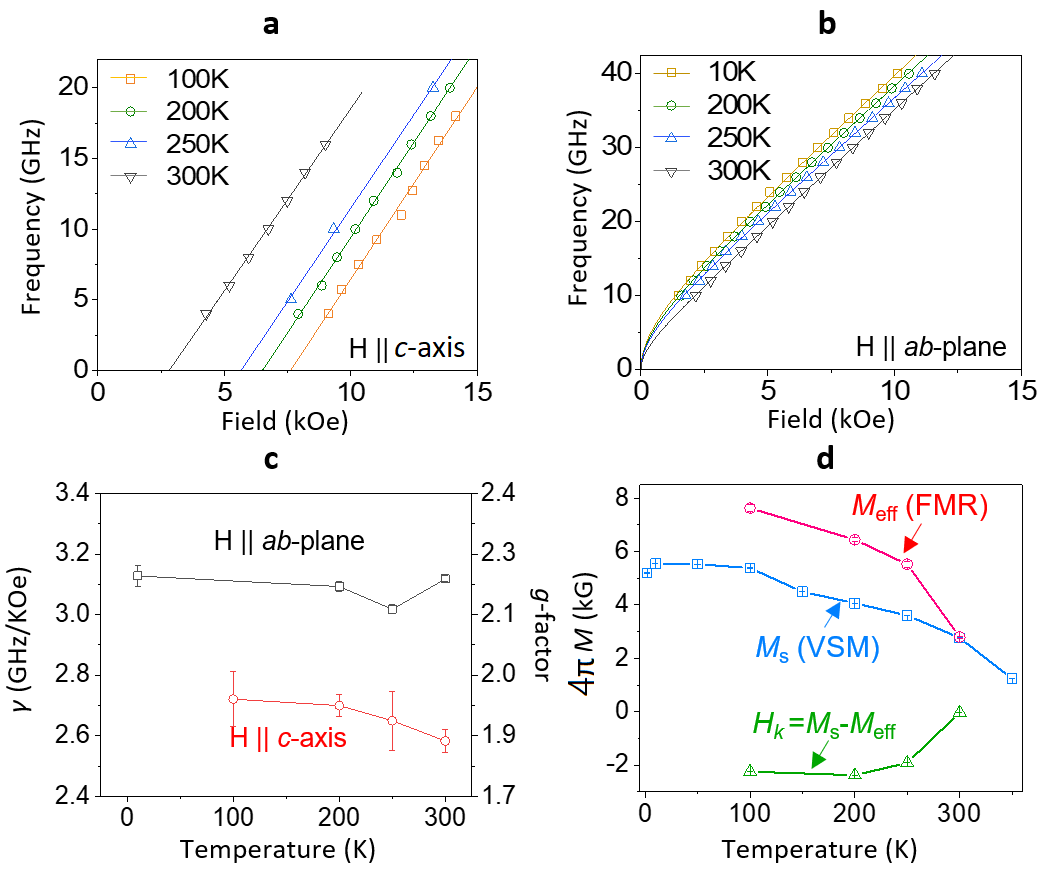}
    \caption{Analysis of the FMR data of the Fe$_5$GeTe$_2$ single crystal. \textbf{a}. Frequency \textit{f} vs. resonance field \textit{H}$_r$ at 100 K, 200 K, 250 K, and 300 K for \textit{H}$\parallel$\textit{c}. The data points are fitted to Eq. (2). \textbf{b}. Frequency \textit{f} vs. resonance field \textit{H}$_r$ at 10 K, 200 K, 250 K, and 300 K for \textit{H}$\parallel$\textit{ab}. The data points are fitted to Eq. (3). \textbf{c}. Temperature dependence of the gyromagnetic ratio $\gamma$ and spectroscopic \textit{g}-factor for the \textit{H}$\parallel$\textit{c} (red) and \textit{H}$\parallel$\textit{ab} (black) cases, respectively. \textbf{d}. Temperature dependence of saturation magnetization 4$\pi$M$_\textrm{s}$ and effective magnetization 4$\pi$M$_\textrm{eff}$ from VSM and FMR measurements, respectively.}
    \label{fig:fig4}
\end{figure}

\begin{figure}[h]
\includegraphics[width=0.95\columnwidth]{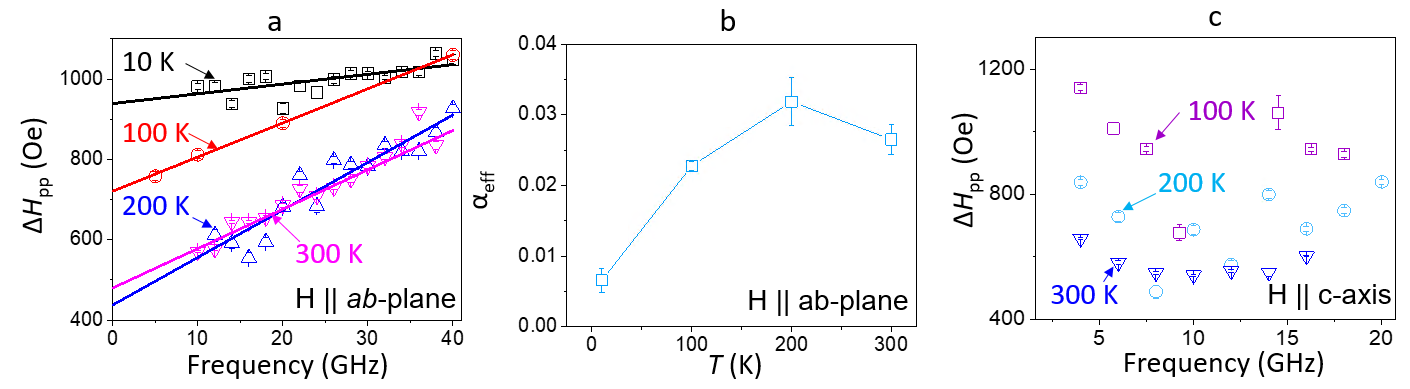}
\caption{\label{fig:py} Characterization of FMR linewidth in Fe$_5$GeTe$_2$. \textbf{a}. Peak-to-peak linewidth $\Delta H_\textrm{pp}$ vs. frequency for \textit{H}$\parallel$\textit{ab}. \textbf{b}. Temperature dependence of effective damping parameter $\alpha_\textrm{eff}$ for \textit{H}$\parallel$\textit{ab}. \textbf{c}. $\Delta H_\textrm{pp}$ vs. frequency for \textit{H}$\parallel$\textit{c}.}
\label{fig:fig5}
\end{figure}

As shown in Figure \ref{fig:fig3}, we detected strong FMR responses at 300 K, demonstrating ferromagnetism of Fe$_5$GeTe$_2$ at room temperature. Figures \ref{fig:fig3}a and \ref{fig:fig3}b show the temperature dependence of the FMR profiles at 10 GHz and 20 GHz, for \textit{H}$\parallel$\textit{c} and \textit{H}$\parallel$\textit{ab}, respectively. Besides the data points, the curves show fits to the derivative of a combination of symmetric and antisymmetric Lorentzian functions\cite{Oates2002}. From the fits one can extract the resonance field \textit{H}$_r$ and peak-to-peak linewidth $\Delta$\textit{H}$_\textrm{pp}$ as shown in Figures \ref{fig:fig4} and \ref{fig:fig5}. It is observed that the magnitude of the FMR peaks decays with reducing temperature along the \textit{H}$\parallel$\textit{c} direction. Below 100 K, the FMR signal becomes undetectable in this orientation. We attribute this phenomenon to the broadening of the FMR resonance peaks. As illustrated in the later discussions in this section, the broadening and weakening FMR peaks indicate the development of an increasing number of magnetic domains with random orientations at lower temperatures.

The resonance frequencies \textit{f} vs. the FMR resonance fields \textit{H}$_\textrm{r}$ at different temperatures are plotted in Figures \ref{fig:fig4}a and \ref{fig:fig4}b for the \textit{H}$\parallel$\textit{c} and for the \textit{H}$\parallel$\textit{ab} cases, respectively. The fitting equation for the \textit{H}$\parallel$\textit{c} measurements is:\cite{LowYIG}
\begin{equation}
     f = \gamma' (H_r -4\pi M_\textrm{eff}) \label{Kittel_OOP}
\end{equation} 
and the fitting equation for the \textit{H}$\parallel$\textit{ab}-plane measurements is:\cite{LowYIG}
\begin{equation}
    f = \gamma' \sqrt{(H_r +4\pi M_\textrm{eff}) H_r}  \label{Kittel_IP}
\end{equation}


\noindent where \textit{f} is frequency, $\gamma'$ is the reduced gyromagnetic ratio ($\gamma' = \frac{|\gamma|}{2\pi}$), and $4\pi$\textit{M}$_\textrm{eff}$ is the effective magnetization. The fitted curves are also presented in Figures 4a and 4b. By constraining the combined fit such that $M_\text{eff}$ is the same along \textit{H}$\parallel$\textit{c} and  \textit{H}$\parallel$\textit{ab}, we obtain different $\gamma'$ and corresponding spectroscopic Landé \textit{g}-factor values for these orientations as shown in Figure \ref{fig:fig4}c. 
\noindent The left vertical axis shows $\gamma'$, and the right vertical axis shows the Landé \textit{g}-factor calculated using $|\gamma|=g\frac{\mu_\textrm{B}}{\hbar}$.  The \textit{g}-factor exhibits a weak dependence on temperature along both the \textit{ab}-plane and the \textit{c}-axis directions. However, it deviates from \textit{g} = 2: it is between 2.05 and 2.23 in the \textit{H}$\parallel$\textit{ab} case, while it is between 1.88 and 1.99 in the \textit{H}$\parallel$\textit{c} case. Thus, our data appears to indicate a sizable difference of the \textit{g}-factor along different directions in Fe$_5$GeTe$_2$.

Similar to Cr$_2$Ge$_2$Te$_6$,\cite{PhysRevB.100.134437} the deviation of the g-factor from \textit{g} = 2 may suggest an orbital contribution to the magnetization due to spin-orbit coupling in Fe$_5$GeTe$_2$. It was found that strong spin-orbit coupling results in nontrival Berry phase in Fe$_3$GeTe$_2$, another member in the Fe$_{n}$GeTe$_2$ (\textit{n}$\geq$ 3) family. In this case, the orbital character is formed by a mixture of 3D orbitals from the Fe I–Fe I dumbbells and Fe II sites \cite{Kim2018}. In Fe$_5$GeTe$_2$, the spin-orbit coupling could be characterized by the \textit{d} orbitals of Fe atoms and \textit{p} orbitals of Te atoms \cite{PhysRevB.100.134437}. In addition, the anisotropy of the g-factor, which follows from that of the orbital moment, is also expected physically: a small orbital moment arising from reduced crystalline symmetry may “lock” the large isotropic spin moment into its favorable lattice orientation through spin-orbit coupling, giving rise to a sizable magnetic anisotropy. Therefore, it is likely that the orbital moment is closely linked to the magnetocrystalline anisotropy in itinerant ferromagnets, as shown theoretically by Bruno et al.\cite{PhysRevB.39.865} for transition-metal monolayers. To form a clearer understanding of the contribution of the orbital momentum, x-ray magnetic circular dichroism (XMCD) measurements are planned for future studies.

However, an unsaturated magnetization at FMR resonance can also lead to an inaccurate estimation of the gyromagnetic ratio.\cite{PhysRevB.96.054436} Because the $\textit{c}$-axis magnetization saturates at significantly large magnetic fields (e.g., H$_\textrm{sat}$ $\approx$ 6 kOe at 100 K), it is possible that the magnetization was not fully saturated at FMR. To estimate to first order the influence of an unsaturated sample at resonance, we write $4\pi M_\textrm{eff}(H) = 4\pi M_\textrm{eff,0} + 4\pi pH$, where $H$ is the external field, $4\pi M_\text{eff,0}$ is the effective magnetization extrapolated to zero field, and $4\pi p$ is the slope of $4\pi M_\textrm{eff}$ vs. $H$ curve in the region where FMR occurs. Using this equation, equation (2) can be written as: $f = \gamma'_\textrm{meas}(H_r-4\pi M_\textrm{eff,meas})$, with $\gamma'_\textrm{meas} = \gamma'(1-4\pi p)$ and $4\pi M_\textrm{eff,meas} = \frac{4\pi M_\textrm{eff}}{1-4\pi p}$. 
Using the VSM data of Figure \ref{fig:fig2}a, we find that the magnetization can change by 10\% over the FMR resonance field range, which can possibly explain the difference of the \textit{g}-factor values measured along the $\textit{c}$-axis and the $\textit{ab}$-plane directions. Thus, we conclude that while the $g$-factor of Fe$_5$GeTe$_2$ may be anisotropic, the broad linewidth and slow saturation for \textit{H}$\parallel$\textit{c} provide another possible explanation. Further clarification could be achieved using high field FMR, but this is beyond the scope of the current manuscript.

The fits also yield the effective magnetization $4\pi$\textit{M}$_\textrm{eff}$ at 100 K, 200 K, 250 K and 300 K. Figure 4d plots $4\pi$\textit{M}$_\textrm{eff}$ and the saturation magnetization $4\pi$\textit{M}$_\textrm{s}$ measured from FMR and VSM, respectively. One can see that there is a difference between $4\pi$\textit{M}$_\textrm{s}$ and $4\pi$\textit{M}$_\textrm{eff}$, revealing an effective anisotropy field. As plotted in Figure 4d, \textit{H}$_\textrm{k}$ = $4\pi$\textit{M}$_\textrm{s}$ - $4\pi$\textit{M}$_\textrm{eff}$ is negative and shows a moderate easy-plane anisotropy field between 100 K and 300 K. The effective anisotropy coefficient \textit{K}$_\textrm{eff}$ is -4.81$\times$10$^5$ erg/cm$^3$ at 100 K and reduces to -3.6$\times$10$^3$ erg/cm$^3$ at 300 K. The negative sign indicates an easy-plane anisotropy. This is consistent with a previous report.\cite{PhysRevB.102.064417}

Next, we analyze the FMR linewidth in order to gain insights on the spin scattering mechanisms in Fe$_5$GeTe$_2$. In Figures \ref{fig:fig5}a and \ref{fig:fig5}c, we plot the peak-to-peak linewidth $\Delta H_\textrm{pp}$ vs. frequency at different temperatures measured for the \textit{H}$\parallel$\textit{ab} and the \textit{H}$\parallel$\textit{c} cases, respectively. For an ideal magnetic thin film that is homogeneous and defect-free, the linewidth vs. frequency plot reflects the intrinsic FMR damping; the Gilbert damping parameter $\alpha$ can be extracted from the slope. $\alpha$ usually increases or decreases with temperature as a consequence of interband or intraband scattering.\cite{StilesJAP} However, due to non-uniform magnetization states and defects in the sample, the linewidth can be broadened by extrinsic scattering mechanisms such as inhomogeneous line broadening $\Delta H_\textrm{0}$ and two-magnon scattering $\Delta H_\textrm{TMS}$. Thus, the FMR linewidth $\Delta H_\textrm{pp}$ can be expressed by the following form:\cite{FMRTMS+IHL}

\begin{equation}
    \Delta H_\textrm{pp}=\frac{2\alpha_\textrm{eff}}{\sqrt{3}|\gamma|}\frac{f}{2\pi}+\Delta H_\textrm{0}+\Delta H_\textrm{TMS}
\end{equation}

Through the analysis of the data in Figure 5, we note our main observations as follows: First, the Fe$_5$GeTe$_2$ linewidth is broadened by extrinsic scattering mechanisms such as $\Delta H_\textrm{0}$ and $\Delta H_\textrm{TMS}$. The increment of $\Delta H_\textrm{0}$ is more pronounced when temperature reduces. Larger $\Delta H_\textrm{0}$ at lower temperatures is a consequence of internal local moment distribution. Because the Fe$_5$GeTe$_2$ sample is single crystal that is defect-free, the larger $H_\textrm{0}$ indicates some intriguing magnetic orders. We will discuss this point further in the discussions below.

Second, as can be seen from Figure 5b, the effective damping parameter $\alpha_\textrm{eff}$ changes between 0.007 and 0.032 from 10 K to 300 K. The magnitude is similar to that of soft 3D transition metal magnets such as Permalloy.\cite{Zhao2016} Because $\alpha_\textrm{eff}$ is estimated from the \textit{H}$\parallel$\textit{ab} measurement, it is likely that $\Delta H_\textrm{TMS}$ also contributes to $\alpha_\textrm{eff}$. In this case, the extracted $\alpha_\textrm{eff}$ provides the upper boundary of intrinsic damping of Fe$_5$GeTe$_2$.

We now summarize all the measured data and provide our perspectives on the intriguing magnetism in Fe$_5$GeTe$_2$: The hysteresis loops show that Fe$_5$GeTe$_2$ is ferromagnetic with an easy-plane magnetic anisotropy at all the temperatures. The \textit{ab}-plane FC curve in Figure 2b shows a sharp drop of magnetization from 200 K to 100 K and flattens upon further reducing $T$. These features indicate that magnetic domains of different orientations were developed with decreasing $T$. The FMR data further support this hypothesis: the linewidth increases as $T$ reduces along both the \textit{ab}-plane and the \textit{c}-axis. The disappearance of the FMR peaks below 100 K along the \textit{c}-axis indicates a higher-level randomness of the magnetic domains compared with the \textit{ab}-plane. The features observed in our work are consistent with those reported in Ref. \cite{PhysRevB.102.064417}, which proposes a complicated magnetism picture. Nevertheless, we show that all the intriguing features in Fe$_5$GeTe$_2$ can be explained by a ferromagnetic phase with randomized magnetic domains at lower $T$.

\section{Summary and conclusion}
In summary, we have synthesized a vdWs Fe$_5$GeTe$_2$ crystal that showed a bulk Curie temperature of 332 K. While the Curie temperature of Fe$_5$GeTe$_2$ is expected to decrease when the 2D vdWs crystal is exfoliated into thin layers, the bulk value is still one of the highest recorded for a 2D magnet until now, making it an attractive 2D option to be used in spintronic devices. We used both VSM and FMR to study the magnetic properties of the Fe$_5$GeTe$_2$ sample. The experiments were performed with external magnetic fields applied along the \textit{c}-axis and the \textit{ab}-plane directions from 300 K to 10 K. We have focused on the temperature and field dependences of the \textit{g}-factor and spin scattering mechanisms. The \textit{g}-factor values are differentiated by a sizable $\Delta$\textit{g} = 0.3 between the \textit{H}$\parallel$\textit{c} and \textit{H}$\parallel$\textit{ab} cases, indicative of considerable orbital momentum arising from spin-orbit coupling in Fe$_5$GeTe$_2$ or fluctuation caused by lack of saturation at FMR. The FMR linewidth analysis reveals low temperature-enhanced inhomogeneous line broadening, together with the VSM data, they indicate a ferromagnetic phase with randomized magnetic domains at lower $T$. For future studies, it will be interesting to exploit Fe$_5$GeTe$_2$ thin films for spin tranport and spin-to-charge conversion experiments at room temperature.



\begin{acknowledgement}
L.B. is supported by the US DOE, Basic Enery Sciences program through award DE-SC0002613. A portion of this work was performed at the National High Magnetic Field Laboratory, which is supported by the National Science Foundation Cooperative Agreement No. DMR-1644779 and the State of Florida. M.B. is supported by the U.S. Department of Energy, Office of Science, Basic Energy Sciences, Materials Sciences and Engineering Division. B.N. acknowledges support through NSF MEMONET grant \#1939999, A.S. acknowledges support NSF-CAREER Award \#1452670, and P.L. acknowledges the Auburn University faculty start-up funding and valuable discussions with Prof. Mingzhong Wu, Prof. Satoru Emori, Prof. Wei Zhang, Dr. Chuanpu Liu, and Dr. Kaya Wei.

\end{acknowledgement}





\bibliography{achemso-demo}

\end{document}